%% file: main.tex
\documentclass[sigplan, nonacm, 10pt]{acmart}
\usepackage{multirow}
\usepackage{listings}

\usepackage{algorithm}
\usepackage{algpseudocode}
\usepackage[normalem]{ulem}
\usepackage{xspace}

\usepackage{caption}
\algrenewcommand\algorithmiccomment[1]{\hfill\textit{// #1}}

\newcommand{\leadpara}[1]{\noindent\textbf{#1}\hspace{.5em}}
\newcommand{\middlepara}[1]{\vspace{.3em}\noindent\textbf{#1}\hspace{.5em}}

\renewcommand\footnotetextcopyrightpermission[1]{}
\newcommand{\sysname}{KVTether}
\newcommand{\syspkg}{kvtether}
\newcommand{\sys}{\sysname\xspace}
\AtBeginDocument{%
  }

\setcopyright{none}

\newcommand{\affmark}[1]{\texorpdfstring{\textsuperscript{#1}}{}}
\newcounter{paperauthorsep}
\makeatletter
\renewcommand{\@mkauthors}{%
  \gdef\@currentauthors{}%
  \global\setbox\mktitle@bx=\vbox{%
    \unvbox\mktitle@bx
    \centering
    \begingroup
      \large\normalfont
      \setcounter{paperauthorsep}{0}%
      \def\and{\stepcounter{paperauthorsep}%
        \ifnum\value{paperauthorsep}=5\par\else\quad\fi}%
      \authors\par
    \endgroup
    \medskip
    {\normalsize\normalfont
      \textsuperscript{1}Hong Kong University of Science and Technology\par
      \textsuperscript{2}Alibaba Group\qquad
      \textsuperscript{3}Shanghai Jiao Tong University\par}
    \smallskip
    {\footnotesize\normalfont\textsuperscript{*}Equal contribution.\par}
    \bigskip}}
\makeatother

\begin{document}

\title[]{Capture the lifecycle: KV Cache management in ReAct Agents with \sys}

\author{Kaihua~Fu\affmark{1,*}}
\author{Yukun~Zhou\affmark{1,*}}
\author{Chaokun~Chang\affmark{1}}
\author{Yinghao~Yu\affmark{2}}
\author{Luping~Wang\affmark{2}}
\author{Guodong~Yang\affmark{2}}
\author{Jiuchen~Shi\affmark{3}}
\author{Quan~Chen\affmark{3}}
\author{Wei~Wang\affmark{1}}

\renewcommand{\shortauthors}{Fu et al.}
\authorsaddresses{}

\input{content/00_abstract}

\maketitle
\hypersetup{pdfauthor={Kaihua Fu, Yukun Zhou, Chaokun Chang, Yinghao Yu,
  Luping Wang, Guodong Yang, Jiuchen Shi, Quan Chen, Wei Wang}}
\pagestyle{plain}
\thispagestyle{plain}

\input{content/01_introduction}
\input{content/03_motivation}
\input{content/03_overview}
\input{content/05_semantic_primitive_model}
\input{content/06_lifecycle_manager}
\input{content/07_cache_engine}
\input{content/08_implementation}
\input{content/09_evaluation}
\input{content/10_conclusion}

\bibliographystyle{ACM-Reference-Format}
\bibliography{sample-base}

\end{document}

%% file: content/00_abstract.tex
\begin{abstract}

Efficient serving of long-context reasoning-and-acting (ReAct) agents relies on KV cache reuse to reduce large language model (LLM) prefill latency and monetary cost. However, a semantic gap exists between agent harnesses and the underlying serving stack.
Through context mutation, tool execution, and subagent coordination, context messages may become actively engaged, permanently discarded, and temporarily unused, while the serving stack only observes accesses to the corresponding KV cache.
This lifecycle blindness prevents recency-only policies such as LRU from reclaiming dead KV promptly and from preserving older KV that will be reused sooner than newer entries.

We present \textit{\sys}, a lifecycle-aware KV cache management framework for ReAct agents. By tracing semantic primitives embedded in agent harnesses, \sys captures runtime lifecycle semantics during highly dynamic execution. \sys then translates message-level semantics into KV-level lifecycle states and uses these states to drive state-prioritized cache management without exposing physical complexities to agent harnesses. After reclaiming dead KV, \sys preferentially preserves live-but-idle KV that is waiting for reuse, reducing premature eviction before reuse. Across agent benchmarks and
production workloads, \sys reduces end-to-end request latency by up to
26.3\% and 17.4\% relative to LMCache and MORI, respectively, and lowers
estimated task cost by 40.0\% and 33.2\% on average.

\end{abstract}

%% file: content/01_introduction.tex
\section{Introduction}
\label{sec:introduction}

Large language model (LLM) applications are evolving from interactive chatbots
to autonomous agents. Mainstream agents such as Claude
Code~\cite{anthropic2025claudecode}, Codex~\cite{openai2025codex},
OpenClaw~\cite{openclaw2026source}, and Hermes~\cite{nous2026hermessource}
execute user requests through a reasoning-and-acting (ReAct) loop~\cite{yao2022react}
interleaving LLM inference with tool calls. Over tens to hundreds of ReAct turns,
a request can accumulate hundreds of thousands of tokens; for example, a Codex
request can reach a 258K-token context. KV cache reuse is essential to avoid
repeated prefill of these long contexts~\cite{choukse2025splitwise}, reducing
inference latency and cost. Yet retaining KV is costly: in
GLM-4.7~\cite{zai2025glm47}, a 258K-token context requires 93GB of KV cache.
Efficient KV-cache management is therefore essential for practical ReAct-agent serving.

\begin{figure}
    \centering
    \includegraphics[width=\linewidth]{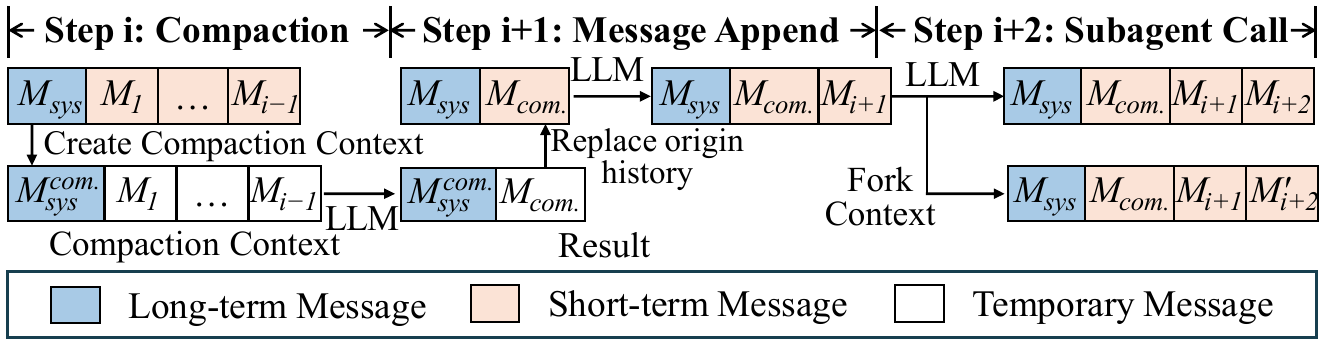}
    \caption{An example of message lifecycle changes across three LLM calls. Within each context, long-term messages are retained for continued reuse across calls; short-term messages remain reusable until removed or replaced; temporary messages expire after the current call.}
    \label{fig:intro}
    \vspace{-3mm}
\end{figure}

ReAct agents, however, expose a fundamental \emph{semantic gap} between the
application layer and the underlying serving stack. At the application layer,
the agent harness manages contexts through message-level operations~\cite{zhang2026context} such as
append, replace, remove, fork, and one-shot construction, which define the
\emph{logical lifecycle} of messages: how long they remain reusable within each
context (Figure~\ref{fig:intro}). The LLM inference
layer (e.g., vLLM~\cite{kwon2023pagedattention} and
SGLang~\cite{zheng2024sglang}) translates these contexts into token sequences,
from which the KV-cache storage layer (e.g., MoonCake
Store~\cite{qin2025mooncake} and LMCache~\cite{cheng2025lmcache}) observes only
cache pages, shared prefixes, and access events. As a result, physical KV
residency can diverge from the logical lifecycles of source messages: the
storage layer cannot tell whether a cached KV span is logically live, explicitly
retained, temporarily waiting, or already dead.

Our analysis of two production traces from a large cloud provider and a
widely used agent benchmark reveals two common inefficiencies caused by
this lack of lifecycle visibility (\S\ref{subsec:root-causes}). First, KV can
remain physically resident as \emph{zombie KV} after its logical lifetime ends.
For example, after a compaction
subagent replaces a long history with a summary, both the discarded history and
the one-shot compaction context are logically dead, yet their KV may persist
until passive eviction. Our replay shows that zombie KV can occupy up to
36.3\% of LMCache's capacity. Second, recency-only policies such as LRU can misorder
live KV. Across the three workloads, 71.0--95.0\% of LRU evictions are
wrong-direction: admitted KV is needed later than the resident KV it displaces.
This happens because recently produced KV in ReAct agents often awaits tool
execution, scheduling, or subagent coordination rather than being closest to
reuse. Both inefficiencies arise because the serving stack cannot distinguish
dead from live KV or waiting from cold KV.

These observations motivate \emph{lifecycle-aware} agent serving, which requires
exposing lifecycle information to the serving stack as it is generated.
Prior works on agentic
workflows ~\cite{zheng2026pbkv, bian2026tokencake} infer KV importance from
\emph{predefined} workflow structures~\cite{openai_deep_research, RAG-Survey, shen2023hugginggpt} or execution histories, but this is
insufficient for ReAct agents. Their execution topology is created at runtime
through context mutations, tool results, and subagent decisions rather than by a
stable structure visible in advance; in one industrial trace, the fraction of
LLM rounds executed by sub-agents ranges from 1.6\% to 86.5\% across requests
(\S\ref{subsec:reuse-dynamics}). More fundamentally, lifecycle-defining events
such as message replacement, removal, one-shot construction, and retention exist
only in the harness’s message-level context management, so lower layers cannot
infer how KV should track application-level context changes or be retained for
future reuse. ReAct-agent serving therefore requires lifecycle events to be
exported from the application layer via runtime tracing and in-harness
integration, rather than inferred downstream.

Making such lifecycle export practical is challenging for three reasons.
\emph{(1) Diverse agent harnesses}. Different agents, and even different
subagents within one agent, may realize message-level context composition
through incremental appends, prompt rebuilding, selective history carry-over, or
one-off contexts, so ad hoc harness-side hooks quickly become brittle and hard
to generalize. Practical lifecycle export therefore requires a harness-side
integration interface that captures common context-management semantics across
diverse harness behaviors. \emph{(2) Cross-layer complexity}. Lifecycle
information originates at the application layer, while KV management operates on
physical objects such as token prefixes and cache pages. The goal is to bridge
these layers without requiring application developers to reason about paging,
prefix organization, or eviction timing; otherwise the lifecycle-export mechanism
merely shifts KV-management complexity upward~\cite{gim2025pie}. \emph{(3)
Uncertain reuse intervals}. Even after lifecycle events are captured and dead KV
can be identified, live-but-idle KV remains hard to manage because their reuse
intervals vary with tool behavior, LLM decisions, and deployment conditions~\cite{chang2026ASBench}.
These dynamics motivate robust KV management under runtime uncertainty,
without relying on precise reuse prediction~\cite{xia2026mori,
li2025continuum, tiwari2026cachewise}.

In this paper, we present \sys, a lifecycle-aware KV-cache management
framework for ReAct agents that exports message-level lifecycle semantics from
the harness to the serving stack. \sys comprises three design modules that
work as a pipeline from semantic export to cache control. \textbf{(1) Semantic
primitive model} (\S\ref{sec:sdk}). \sys places lifecycle export at the
message-operation level, which matches how agent harnesses already compose and
revise contexts. Developers use a small set of \emph{message-level primitives}
directly in existing context-management code, and \sys captures lifecycle
events automatically from those operations. These exported events provide a
common semantic interface across diverse harness behaviors and serve as the
input to \sys's lifecycle manager. \textbf{(2) Event-driven lifecycle
manager} (\S\ref{sec:control-plane}). \sys uses an event-driven manager to
translate exported message-level events into KV lifecycle states as cache action
signals. It bridges the gap between message lists at the application layer and
tokenized prefixes and cache pages in the serving stack. This design preserves
precise KV control while keeping physical details transparent to the application
layer. \textbf{(3) State-guided KV cache engine} (\S\ref{sec:data-plane}).
Guided by the lifecycle states produced by the manager, \sys's KV cache
engine first reclaims space from dead KV, thereby confining the remaining
uncertainty to live-but-idle KV. The further optimization is built upon a key
observation that, when limited cache capacity forces space competition, newly
used KV is often less urgent to retain than older live KV, as the execution of
tools or subagents is not instantaneous. \sys exploits this directional bias
with MRU (most-recently-used) as a straightforward realization, mitigating wrong-direction evictions
without relying on precise reuse prediction.

We evaluate \sys on agent benchmarks and production traces from a large
cloud provider, using 12 Nvidia GB200 GPUs across three nodes
(\S\ref{sec:evaluation}). We compare against LMCache~\cite{cheng2025lmcache},
a strong semantic-blind KV cache, and MORI~\cite{xia2026mori}, an agent-aware
baseline that predicts reuse from inferred request idleness.
Relative to LMCache and MORI, \sys lowers estimated task pricing by
40.0\% and 33.2\% on average and end-to-end request latency by up to 26.3\%
and 17.4\%, respectively. To assess the semantic primitive model's ease of
adoption, we integrate \sys into six agent harnesses with only 26--216
lines of integration code per harness.

%% file: content/03_motivation.tex
\section{Background and Motivation}
\label{sec:background_and_related_work}

In this seciton, we introduce ReAct execution and context lifecycles
(\S\ref{subsec:agent-workflows}), then use workload traces to characterize
KV reuse dynamics (\S\ref{subsec:reuse-dynamics}). We investigate  missed reuse
under existing policies, and show the
resulting prefill cost
(\S\ref{subsec:black-box-cache}). We analyze two underlying cache-management
failures (\S\ref{subsec:root-causes}) and discuss related work
(\S\ref{subsec:related-work-background}).

\subsection{ReAct Execution and Context Lifecycles}
\label{subsec:agent-workflows}

\leadpara{ReAct execution and context messages.}
Reasoning-and-acting (ReAct)
agents~\cite{anthropic2025claudecode,openai2025codex,openclaw2026source,nous2026hermessource}
execute a user request as a loop that alternates LLM calls with tool
execution~\cite{yao2022react}. A context manager keeps the evolving task state as an ordered
sequence of messages and assembles them into the prompt for each LLM call.
Each message carries a role: \textit{System messages} encode persistent
instructions or memory, \textit{User messages} describe goals and constraints,
and \textit{Assistant} and \textit{Tool messages} record model outputs and tool
results.

\begin{figure}[t]
  \centering
  \includegraphics[width=0.95\linewidth]{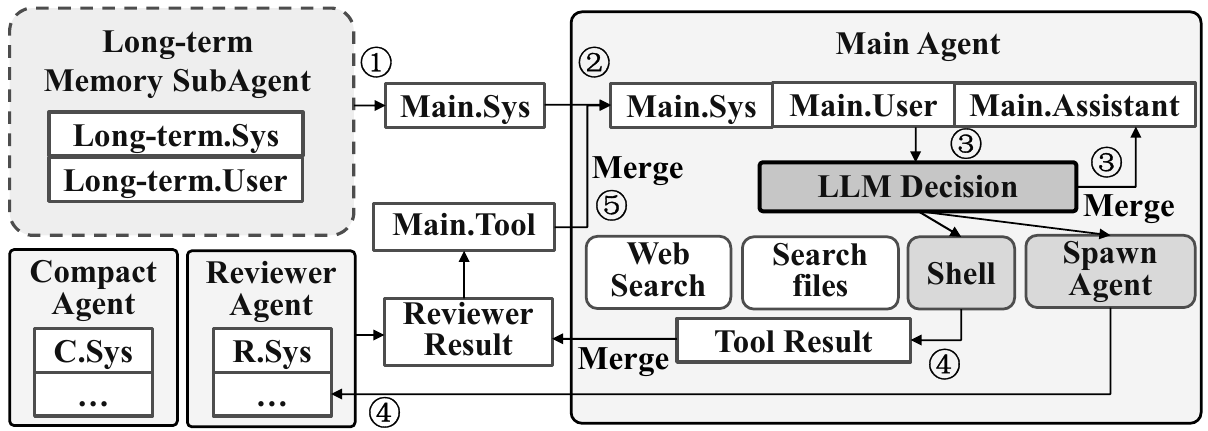}
  \caption{A ReAct agent repeatedly composes message contexts, calls the LLM, waits for tools or sub-agents, and updates the context for later rounds.}
  \label{fig:react-agent-workflow}
\end{figure}

Figure~\ref{fig:react-agent-workflow} traces one such loop.
\textcircled{1} A long-term-memory sub-agent prepares the System message.
\textcircled{2} The main agent initializes its context with System and User messages.
\textcircled{3} The LLM selects tools or sub-agent tasks and records its decisions as
Assistant messages.
\textcircled{4} The agent issues these calls and waits for the results.
\textcircled{5} The context manager appends the results as messages.
Steps \textcircled{3} to \textcircled{5} repeat until the task finishes, and sub-agents
keep their own contexts and may run in parallel. Because each tool and
sub-agent choice depends on the accumulated context, the execution topology
emerges at runtime rather than following a predefined workflow.

\middlepara{Message and context lifecycles.}
This dynamic execution creates diverse message lifecycles.
Main-agent history may be reused across rounds, while a reviewer sub-agent
may use task-specific messages for a single inference. A compaction sub-agent
replaces older messages in the main context with a summary, ending their
reuse while retaining selected instructions or memory. Child contexts may
inherit parent prefixes yet evolve independently, so removing a message or
terminating a context need not end reuse elsewhere. These operations determine
whether the corresponding KV remains logically reusable.

\subsection{KV Reuse under Dynamic Agent Execution}
\label{subsec:reuse-dynamics}

We characterize how these execution and lifecycle dynamics shape KV reuse
across three ReAct-agent workloads.
The \uline{\emph{Codex}} workload uses a self-deployed software engineering
agent based on Codex~\cite{openai2025codex} to run
TerminalBench~\cite{merrill2026terminalbench}. \uline{\emph{Agent-X}} and
\uline{\emph{Agent-G}} are two production coding agents deployed by a large
cloud provider, which have a combined peak of over 27,000
active sessions; we collect 400 trajectories per agent.
Table~\ref{tab:agent-dynamics} summarizes per-request statistics.

\begin{table}[t]
  \centering
  \caption{Min--max ranges of per-request characteristics across Codex and two production ReAct-agent traces. KV size is computed from total prefill tokens for GLM-4.7~\cite{zai2025glm47}.}
  \vspace{-2mm}
  \footnotesize
  \begin{tabular}{c|c|c|c}
    \hline
    \textbf{Metric} & \textbf{Agent-G} & \textbf{Codex} & \textbf{Agent-X} \\ \hline
    Total LLM rounds & 9--106 & 3--97 & 3--127 \\
    \hline
    Total prefill tokens & 17K--312K & 9K--363K & 11K--374K \\
    \hline
    Total decode tokens & 2K--16K & 536--13K & 2K--11K \\
    \hline
    Total KV size (GB) & 6--112 & 3--130 & 4--134 \\
    \hline
    Sub-agent LLM round ratio & 2.8\%--25.0\% & 2.0\%--17.7\% & 1.6\%--86.5\% \\
    \hline
    Distinct tool-call types & 3--21 & 2--47 & 1--35 \\
    \hline
  \end{tabular}
  \label{tab:agent-dynamics}
\end{table}

\middlepara{Large reusable contexts.}
As messages accumulate across rounds, successive LLM calls often share
long prefixes. Requests span 3--127 LLM rounds and accumulate 9K--374K prefill
tokens (Table~\ref{tab:agent-dynamics}), making repeated prefill of these prefixes
costly. KV reuse avoids this computation, but retaining reusable prefixes
consumes cache capacity: these token counts correspond to 3--134~GB of KV
for GLM-4.7.

\middlepara{Lifecycle behavior varies across agents.}
Sub-agents account for up to 86.5\% of LLM rounds in Agent-X, versus
17.7\% in Codex and 25.0\% in Agent-G. Some sub-agents reuse context across
rounds, so their KV may remain useful for later calls. Others discard one-off
contexts after a single call, ending reuse of their exclusive KV. Token overlap
identifies shared prefixes, but does not reveal when a context ends or whether
another context still needs its KV.

\begin{figure}[t]
  \centering
  \includegraphics[width=0.95\linewidth]{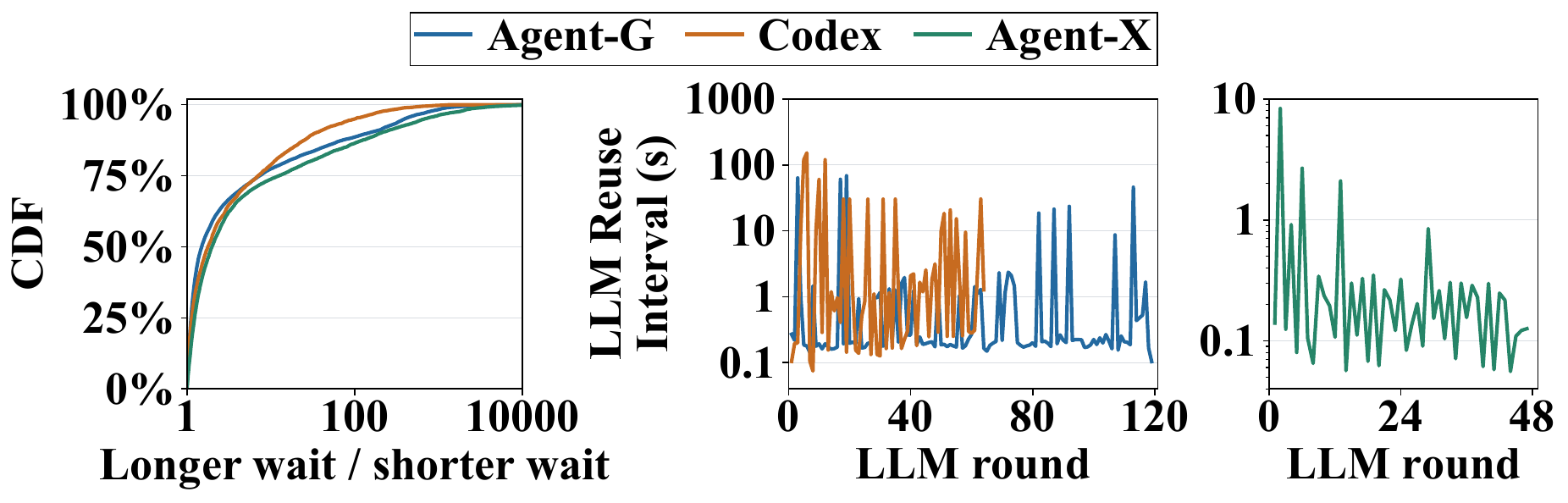}
  \caption{Agent workloads exhibit unstable non-LLM waiting intervals. (a) CDF of the longer-to-shorter ratio for adjacent non-LLM waiting intervals. (b)--(c) Per-round non-LLM waiting times of one representative request per workload.}
  \label{fig:motivation-tool-call-times}
\end{figure}

\middlepara{Reuse intervals are unstable.}
Requests use 1--47 distinct tool types (Table~\ref{tab:agent-dynamics}),
whose execution affects when the next LLM call can reuse the context's live KV.
Other runtime factors, including decoding, sub-agent execution, queueing,
and scheduling, further shape this reuse interval.
Figure~\ref{fig:motivation-tool-call-times}(a) shows wide variation in
consecutive non-LLM waiting intervals. The 90th-percentile longer-to-shorter
ratio is 144.52$\times$, 33.67$\times$, and 204.43$\times$ for Agent-G, Codex, and
Agent-X, respectively.
Panels (b)--(c) also show little temporal continuity within representative
requests.

KV liveness and reuse ordering depend on runtime execution. This motivates
adapting retention decisions online to changing contexts and reuse intervals.

\subsection{Low KV Hit Rates under Semantic-Blind Policies}
\label{subsec:black-box-cache}
\begingroup
\emergencystretch=1em

We now quantify the reuse opportunities missed by existing KV-management
policies under these dynamics.
We replay the three traces on a three-node serving cluster, with the setup
given in Table~\ref{tab:motivation-setup} in \S\ref{subsec:setup}.

We compare two representative baselines, \textbf{LMCache}~\cite{cheng2025lmcache}
(LMC) and \textbf{MORI}~\cite{xia2026mori}, both using a decoupled three-layer
serving stack.
The harness manages messages and assembles prompts for LLM calls.
SGLang~\cite{zheng2024sglang} tokenizes prompts, performs prefill and decode,
and indexes shared prefixes, while LMCache stores the resulting KV.
For KV eviction, LMC uses LRU, whereas MORI ranks KV
by relative idleness using recent LLM-active time and non-LLM waiting time,
including tool and sub-agent waits. We report one representative load point,
using 0.06, 0.05, and 0.06 requests per second (RPS)
for the three workloads; \S\ref{sec:evaluation} studies load sensitivity.

\begin{figure}[t]
  \centering
  \includegraphics[width=\columnwidth]{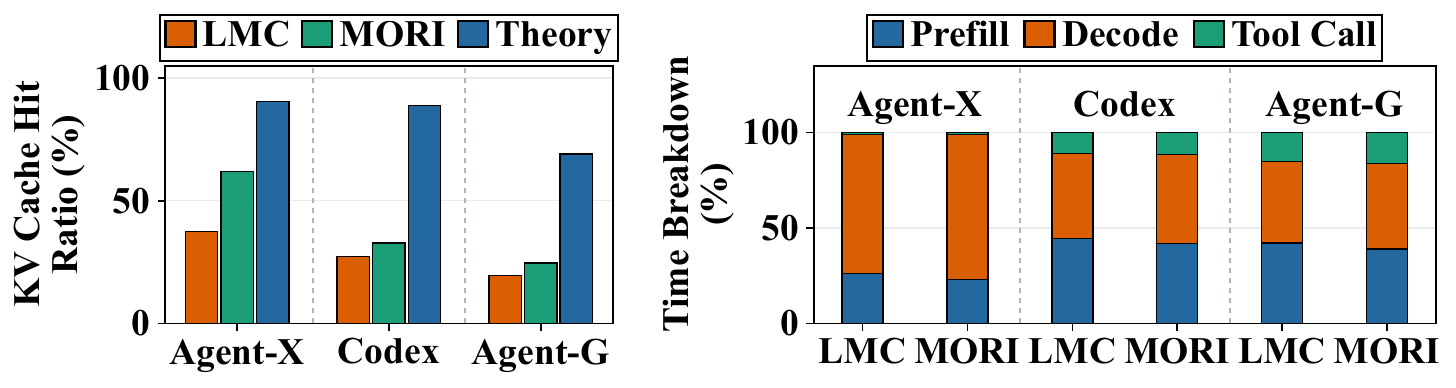}
  \caption{The theoretical maximum versus realized KV cache hit ratios (left) and aggregate execution time breakdown (right) under LMC and MORI on three workloads.}
  \label{fig:motivation-latency-reuse}
\end{figure}

Figure~\ref{fig:motivation-latency-reuse} (left) shows that realized KV cache
hit ratios average only 28.1\% under LMC and 39.8\% under MORI across the three
workloads. Both fall far below the average theoretical maximum of 82.8\%,
which is measured by replaying each workload's requests offline, one at a time.

This decoupled serving stack creates a critical semantic gap between the
harness and KV management. Message lifecycle changes in the harness are not
propagated to lower layers, which manage physical residency per token or page.
Without lifecycle semantics, they fall back on access history and timing signals
to decide which KV to retain under capacity pressure. This can leave obsolete
KV resident while displacing prefixes needed by later calls.

The resulting misses require recomputing reusable prefixes, increasing prefill
work and inference cost. Figure~\ref{fig:motivation-latency-reuse} (right) shows
the breakdown of execution time summed across agents, sub-agents, and tool
calls, including concurrent work. Prefill remains a substantial component,
accounting for 37.6\% under LMC and 34.6\% under MORI on average.
Improving KV cache hit ratios could therefore substantially reduce end-to-end
latency by avoiding repeated prefill.

\subsection{Two Failures Caused by the Semantic Gap}
\label{subsec:root-causes}

This semantic gap leads to two cache-management failures. First, a
\emph{liveness failure} occurs when the system retains KV whose logical
lifetime has ended. Second, an \emph{ordering failure} arises when access
history and timing signals misrepresent future reuse, causing misordered
evictions among live KV.

\middlepara{Liveness failure: zombie residency after lifecycle termination.}
A zombie page is KV whose source message or context has lost all
future reuse value because a request finished, a compaction replaced a
history span, or a one-off context was discarded.
Figure~\ref{fig:motivation-zombie-pages} reports the zombie share in the KV
pool. On Agent-X, Codex, and Agent-G, LMC zombie shares peak at
31.1\%, 36.3\%, and 15.8\%, respectively, because LMC observes
only recency, not lifecycle ends. MORI exhibits higher average zombie
occupancy, with shares peaking at 49.7\%, 33.1\%, and 27.4\%, respectively,
because request idleness can prioritize dead KV indefinitely.

\begin{figure}[t]
  \centering
  \includegraphics[width=\linewidth]{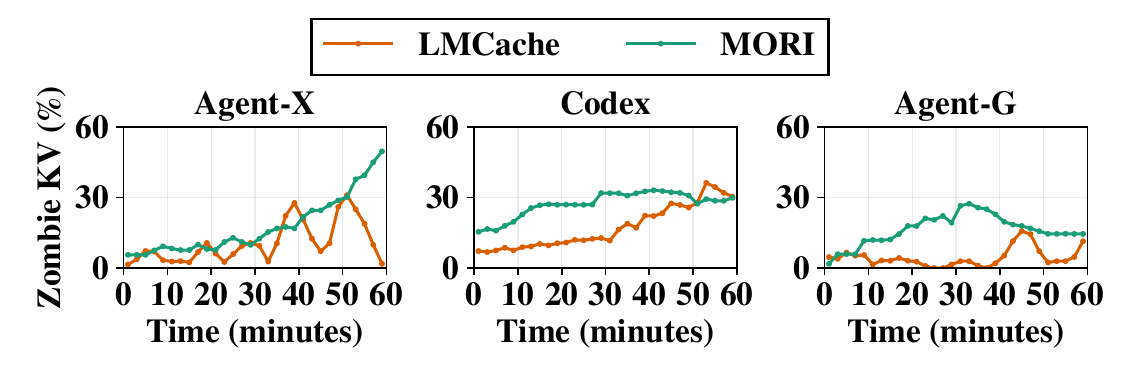}
  \caption{Share of zombie pages in the KV pool under LMC and MORI on the three workloads.}
  \label{fig:motivation-zombie-pages}
\end{figure}

This failure is not fundamentally a prediction problem. Once the harness
removes, replaces, or retires a message span, its exclusive KV has already
lost future reuse value. Failing to reclaim that KV is not due to uncertain
timing, but to missing lifecycle information at the point where logical
reachability changes.

\middlepara{Ordering failure: wrong-direction evictions among live KV.}
Even after excluding dead KV, capacity pressure remains among live-but-idle
KV. Figure~\ref{fig:motivation-wrong-evictions} reports wrong-direction
evictions over one hour on each trace. Following Belady's
rule~\cite{belady1966study}, the best victim is the page whose next use lies
farthest in the future; an eviction is wrong-direction when a newly inserted
page A displaces a resident page B even though A will be reused later than B.

LMC shows average wrong-direction ratios of 79.5\%, 95.0\%, and 71.0\%
on Agent-X, Codex, and Agent-G, respectively, because LRU assumes that
recently used KV is closest to reuse. ReAct agents often invert
this order: newly produced KV frequently enters a waiting period while the
agent decodes, runs tools, executes sub-agents, or waits behind other requests.
MORI lowers these ratios to 57.8\%, 75.6\%, and 45.7\%, respectively, by using
idleness as a reuse proxy, but this still leaves a
substantial fraction of misordered evictions. MORI accounts for observed
idleness, but variable tool and sub-agent delays can decouple this signal
from the remaining time to KV reuse.

\begin{figure}[t]
  \centering
  \includegraphics[width=\linewidth]{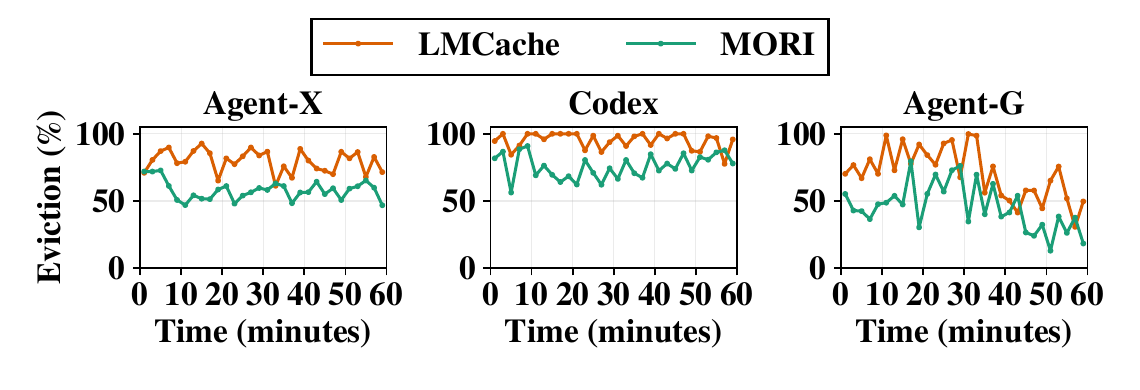}
  \caption{Wrong-direction eviction ratio of LMC and MORI on the three workloads.}
  \label{fig:motivation-wrong-evictions}
\end{figure}

This residual problem remains difficult to resolve through precise prediction
alone. As \S\ref{subsec:reuse-dynamics} shows, decoding, tools,
sub-agents, queueing, and scheduling shape reuse intervals online, causing
wide variation across consecutive rounds. Even for live KV, exact next-use
times remain uncertain. Prediction-based policies may help, but cannot
replace lifecycle semantics for determining liveness or reliably eliminate
ordering errors under this uncertainty.

\subsection{Related Work}
\label{subsec:related-work-background}

Based on these observations, efficient agent KV management requires three
capabilities: tracking harness-side message lifecycles, propagating lifecycle
information across layers for KV retention and reclamation, and ordering
live KV under uncertain reuse intervals. Existing work addresses parts of this space,
but not this harness-to-cache lifecycle full path.

\middlepara{Semantic-blind KV substrates.}
PagedAttention~\cite{kwon2023pagedattention},
RadixAttention~\cite{zheng2024sglang}, LMCache~\cite{cheng2025lmcache},
MoonCake~\cite{qin2025mooncake}, and CachedAttention~\cite{CachedAttention}
provide the physical backbone of KV cache management through paging, prefix indexing,
and tiered KV storage. However, besides the physical token streams,
shared prefixes, and page accesses, they are totally blind to the logical message operations that determine
KV cache reuse in ReAct contexts. Therefore, they only count on conventional recency policies
such LRU for eviction ordering, underlying the two failures in \S2.4.

Semantic-aware serving improves scheduling and execution
coordination~\cite{tan2024teola,lin2024parrot,luo2026agentix, kang2026thunderagent},
motivating similar support for agent KV management. Existing systems
expose TTLs and KV retention ranges~\cite{anthropic2026promptcaching,
openai2026promptcaching,google2026contextcaching,nvidia2026dynamoagenthints},
while Pie~\cite{gim2025pie} exposes fine-grained programmable inference
operations. However, these controls lack a common semantic contract
mapping message operations to KV lifecycle updates. Applications still
translate replacement, removal, fork inheritance, and one-off execution
into backend-specific hints or operations; physical interfaces additionally
expose token/page boundaries, prefix dependencies, and shared-KV
complexities. Lifecycle translation thus remains application-specific.

\middlepara{Workflow-based KV Management.}
TokenCake~\cite{bian2026tokencake}, PBKV~\cite{zheng2026pbkv},
and KVFlow~\cite{KVFlow} use predefined workflow stages and dependencies
to anticipate reuse and guide KV retention and placement in agentic
workflows~\cite{openai_deep_research,RAG-Survey,shen2023hugginggpt},
such as deep research. However, these methods cannot be directly
applied to ReAct agents, whose tool calls and subagent branches emerge
incrementally from LLM decisions and tool results. Execution paths
cannot be exhaustively enumerated in advance, and future topology
remains unknown at cache decision time. The ReAct loop structure
therefore cannot reveal which contexts will be revisited or their
reuse order.

\middlepara{Prediction-based KV policies.}
Continuum~\cite{li2025continuum}, CacheWise~\cite{tiwari2026cachewise},
and MORI~\cite{xia2026mori} use tool-call timing to estimate KV reuse
under dynamic ReAct execution. Continuum and CacheWise derive per-entry
TTLs from historical tool-call interval distributions, while MORI infers
reuse priorities from the fraction of recent execution time spent in
tool calls. These approaches have two limitations. First, without
lifecycle distinctions, dead KV can occupy capacity and displace live
KV, undermining eviction effectiveness. Second, runtime-dependent
execution and changing deployment conditions make reuse intervals
unstable, weakening prediction reliability across workloads and
propagating per-entry uncertainty into eviction decisions.

\par\endgroup

%% file: content/03_overview.tex
\section{Challenges and Design Overview}
\label{sec:overview}

\begin{figure}[t]
    \centering
    \includegraphics[width=0.95\linewidth]{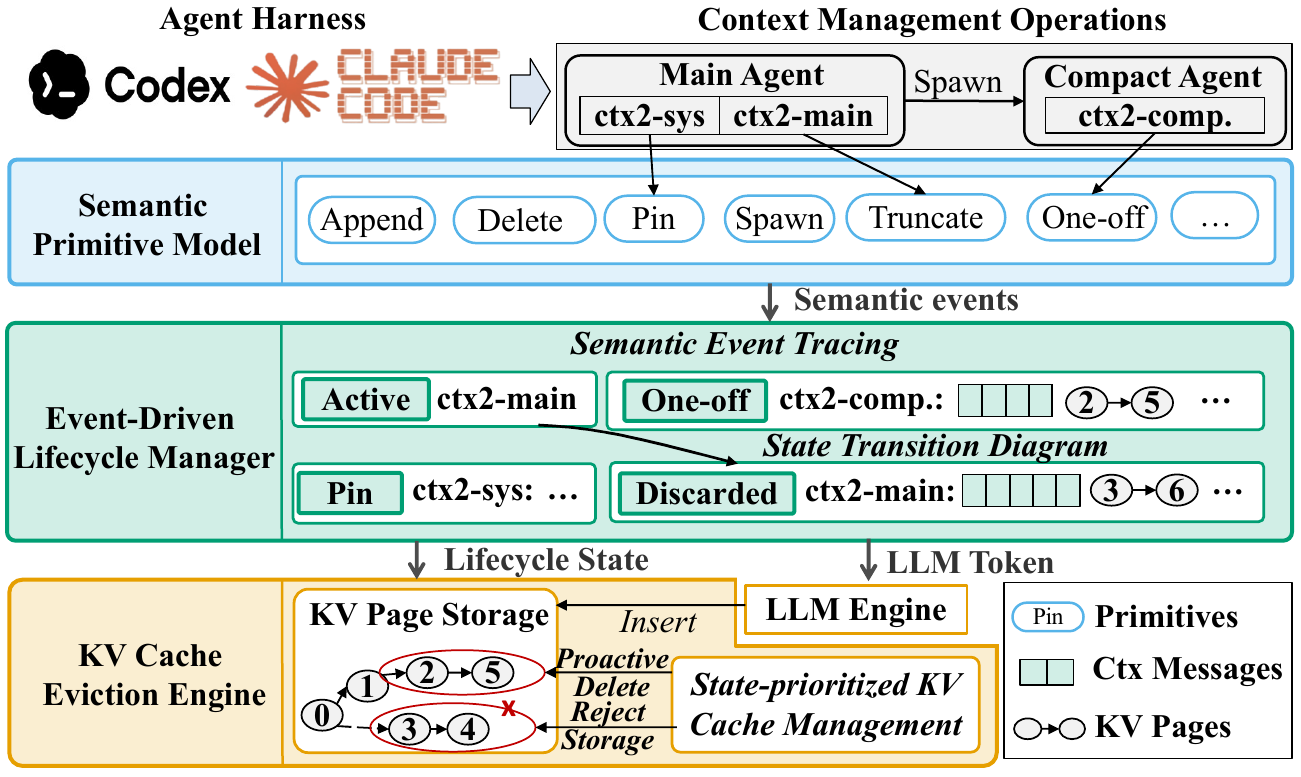}
    \caption{The design overview of \sys.}
    \label{fig:overview}
\end{figure}

The analysis in \S2 shows that the semantic gap in ReAct agents cannot be closed by downstream inference alone: lifecycle semantics should be exported from the harness at runtime.  Making this practical, however, still involves three concrete challenges. \sys follows the architectural lifecycle-export idea and overcomes the challenges through three corresponding designs, summarized in Figure~\ref{fig:overview}.

\middlepara{C1: Diverse agent harnesses.}
Agent harnesses realize context evolution through different context-management patterns, making ad hoc hooks brittle and hard to generalize. Rather than tie lifecycle export to harness-specific functions, \sys targets the more stable substrate that these harnesses share: message-level context operations. It therefore provides a \textit{\uline{semantic primitive model}} that exports lifecycle semantics through a small set of primitives for message edits, context creation, and lifetime declaration. This gives heterogeneous harnesses a common runtime interface for lifecycle export while preserving their existing agent logic.

\middlepara{C2: Cross-layer complexity.}
Lifecycle semantics are generated by message-level context management, while actionable application requires resolving physical details beyond harness-level concerns, including how messages materialize into tokens and KV, how edits invalidate KV suffixes due to prefix dependency, and how multiple contexts deal with conflict message spans on shared prefixes.
\sys addresses this challenge with an \textit{\uline{event-driven lifecycle manager}} that translates message-level events into KV-level lifecycle states. This manager encapsulates tokenization, prefix structure, and sharing details, so the serving stack can apply lifecycle-aware KV control precisely without pushing physical cache-management complexity back to agent harnesses.

\middlepara{C3: Uncertain reuse intervals.}
Even after lifecycle semantics identify dead KV, capacity pressure remains among live-but-idle KV whose reuse intervals vary with tools, sub-agents, and runtime delays. \sys's \emph{\uline{state-guided KV cache engine}} therefore takes a state-prioritized approach. It first bypasses or proactively reclaims dead KV, confining uncertainty to the live-but-idle tier. \sys then orders and evicts KV within this tier using observed overall reuse trends, avoiding depending on precise per-item prediction.

Figure~\ref{fig:overview} also illustrates the end-to-end workflow. As the harness updates a context, it emits lifecycle events through the primitive interface. The lifecycle manager resolves each event to the affected token/KV scope and updates the corresponding KV lifecycle state. Guided by that state, the cache engine turns lifecycle state into cache actions such as bypass, reclamation, and eviction.

%% file: content/05_semantic_primitive_model.tex
\section{Semantic Primitive Model}
\label{sec:sdk}

In this section, we introduce the semantic programming model in \sys,
which exposes context-management intent through message-level primitives.
Developers declare context updates and message lifetimes through interfaces
compatible with their existing harnesses. These operations generate semantic
events for the lifecycle manager (\S\ref{sec:control-plane}).

\subsection{Message-Level Primitives}
\label{subsec:context-abstraction}
\label{subsec:sdk-primitives}
\label{subsec:semantic-classes}

To identify context lifecycle transitions, an intuitive approach is to
classify subagents by function (e.g. memory compaction or tool safety
guardian) and map function invocations to the corresponding transitions.
However, it is challenging to formulate function-based lifecycle rules
that generalize across agent harnesses.
We investigate four agent harnesses (Table~\ref{tab:agent-message-sequences})
implemented in different programming languages. These agents support
diverse functions, and even the same function can have different context
requirements across harnesses.
For example, Codex's safety guardian function~\cite{openai2026codexsource}
retains interaction history, including prior tool commands, across reviews,
whereas Hermes~\cite{nous2026hermessource} assesses the current command without prior history, and SWE-agent~\cite{sweagent2026source} and Agent-G
invoke no LLM.

Despite these differences, we find that all four harnesses rely on
context-management logic that maintains messages and assembles LLM inputs.
We refer to this
abstraction as the context manager. Table~\ref{tab:agent-message-sequences}
shows that all four harnesses store interaction histories as ordered
message sequences, providing an opportunity to track context lifecycle
transitions at message granularity. Tracking message updates captures
context changes, while explicit lifetime declarations capture retention
intent.

\begin{table}[t]
  \centering
  \caption{The Data structure of context managements in agent harnesses.}
  \vspace{-2mm}
  \label{tab:agent-message-sequences}
  \footnotesize
  \setlength{\tabcolsep}{3pt}
  \begin{tabular}{@{}p{0.23\columnwidth}|p{0.15\columnwidth}|p{0.56\columnwidth}@{}}
    \toprule
    \textbf{Agent} & \textbf{Language} & \textbf{Message sequence} \\
    \midrule
    Codex CLI~\cite{openai2026codexsource} & Rust
      & \texttt{Vec<ResponseItemEnvelope>} \\
    SWE-agent~\cite{sweagent2026source} & Python
      & \texttt{list[HistoryItem]} \\
    Hermes~\cite{nous2026hermessource} & Python
      & \texttt{List[\{role: ..., content: ...\}]} \\
    Agent-G & Go
      & \texttt{[]*types.ChatMessage} \\
    \bottomrule
  \end{tabular}
\end{table}

\begin{table}[t]
  \centering
  \caption{Primitives in \sys's library for managing agent contexts and declaring message lifetimes.}
  \vspace{-2mm}
  \label{tab:sdk-primitives}
  \footnotesize
  \setlength{\tabcolsep}{3pt}
  \begin{tabular}{@{}p{0.22\columnwidth}|p{0.17\columnwidth}|p{0.55\columnwidth}@{}}
    \toprule
    \textbf{Category} & \textbf{Primitive} & \textbf{Description} \\
    \midrule
    \multirow[c]{3}{=}{Message edits}
      & \texttt{append}  & Record message appends. \\
      & \texttt{replace} & Record in-place message replacement. \\
      & \texttt{remove}  & Record message removals. \\
    \midrule
    \multirow[t]{2}{=}{Context creation}
      & \texttt{fork}    & Register a child sharing a parent prefix. \\
      & \texttt{spawn}   & Register an independent child context. \\
    \midrule
    \multirow[t]{3}{=}{Lifetime declarations}
      & \texttt{one\_off} & Mark a context for one inference. \\
      & \texttt{pin}     & Declare long-term message retention. \\
      & \texttt{unpin}   & Release retention protection. \\
    \bottomrule
  \end{tabular}
\end{table}

Based on this observation, \sys exposes eight primitives that record
context updates and retention intent for message-level lifecycle tracking
as shown in Table~\ref{tab:sdk-primitives}. Message edits (\texttt{append},
\texttt{replace}, and \texttt{remove}) identify the affected context and
message range; context creation (\texttt{fork} and \texttt{spawn}) records
whether a child inherits a parent prefix. Lifetime declarations
(\texttt{one\_off} and \texttt{pin}) distinguish ordinary reuse from
single-use and explicitly retained messages; \texttt{unpin} releases this protection.

\middlepara{Expressiveness of the Primitive Set.} These primitives capture both
the spatial evolution of contexts and the temporal requirements for message
retention. Spatially,
changes to a message list reduce to adding, modifying, or removing messages;
combining these edits also handles changes in order. Temporally, a new
context either starts independently or inherits existing messages, which
can have different lifetimes in different contexts. Messages may be used
for one inference, reused across rounds, or explicitly retained, while
runtime events mark when their use pauses, resumes, or ends
(\S\ref{sec:control-plane}). These cases cover the context changes and
retention requirements in our model, so more complex functions can be
expressed by combining the existing primitives.

\subsection{Integration and Composition}
\label{subsec:high-level-apis}

Figure~\ref{fig:sdk-integration} combines a Python adapter that subclasses
an existing \texttt{ContextManager} with a context compaction example.
The adapter preserves the harness's message updates and prompt construction,
reports operations through the SDK, and pins System messages during
initialization. The agent harness continues to use the manager without
handling the context's tokens or KV pages. Compaction derives a temporary
context, declares it one-off before inference, and replaces the selected
history with the resulting summary. System messages remain outside the
replaced range, and subsequent rounds reuse the summary. Such sequences can
be encapsulated in higher-level methods while preserving the same semantic
events.

\begin{figure}[t]
\newcommand{\pythonpkg}{\textcolor{violet!75!black}{\syspkg}}
\begin{lstlisting}[language=Python,
  basicstyle=\ttfamily\scriptsize,
  keywordstyle=\color{blue!65!black}\bfseries,
  commentstyle=\color{green!40!black},
  stringstyle=\color{red!55!black},
  emph={pin,append,replace,fork,one_off},
  emphstyle=\color{violet!75!black},
  columns=fullflexible,keepspaces=true,
  showstringspaces=false,breaklines=true,
  numbers=left,numberstyle=\tiny\color{black!45},numbersep=5pt,
  xleftmargin=12pt,frame=tb,
  aboveskip=0pt,belowskip=0pt]
import (*@\pythonpkg@*)
from agent_harness import ContextManager

class (*@\sysname@*)ContextManager(ContextManager):
    def __init__(self, messages):
        super().__init__(messages=messages)
        (*@\pythonpkg@*).pin(messages)

    def append(self, message):
        result = super().append(message)
        (*@\pythonpkg@*).append(self, message)
        return result

    def replace(self, start, end, message):
        result = super().replace(start, end, message)
        (*@\pythonpkg@*).replace(self, start, end, message)
        return result

    def fork(self, end):
        child = super().fork(end)  # Same type; no init.
        (*@\pythonpkg@*).fork(self, child, end)
        return child

context = (*@\sysname@*)ContextManager([system_message])
context.append(user_message)
...  # Agent execution accumulates history.
compaction = context.fork(end)
(*@\pythonpkg@*).one_off(compaction)
compaction.append(summary_instruction)
context.replace(start, end, run_llm(compaction))
\end{lstlisting}
\caption{Simplified SDK integration and context compaction.}
\label{fig:sdk-integration}
\end{figure}

Each derived context has an independent message container. Its termination
ends only its own unpinned references; other contexts may still require the shared
prefix. Similarly, replacing a message range preserves the order of
unaffected messages and changes only the issuing context. The lifecycle
manager reconciles these logical events with shared physical KV ownership.

%% file: content/06_lifecycle_manager.tex
\section{Event-Driven Lifecycle Management}
\label{sec:control-plane}

The primitive model helps expose context evolution as harness-side semantic events, but these events still do not directly specify how physical KV should be managed. The lifecycle manager closes this gap by interpreting them against the physical organization of KV, including message-to-KV mapping, prefix-sensitive validity, and cross-context sharing. It therefore proceeds in two steps: \S\ref{subsec:tag-tracing} resolves the affected KV scope, and \S\ref{subsec:state-automaton} defines the resulting lifecycle states.

\subsection{KV Scope Resolution}
\label{subsec:tag-tracing}
\sys resolves semantic events with context as the primary management unit. For each independently evolving context, the lifecycle manager maintains the mapping from its messages to the materialized token and KV ranges they currently cover. Context-scoped events update an entire context scope directly, while message-scoped events require translation.

This translation in \sys follows the same prompt construction path in LLM engines. Message lists are first formatted into the actual prompt string and then tokenized into the token sequence indexing KV cache. Most events then translate by position, but remove and replace on messages are constrained by \emph{prefix-dependent validity}, where changing a message in the middle of a context invalidates the cached suffix due to the KV cache computing logic in LLM inference, even when later messages are unchanged. \sys therefore translates them into suffix invalidation, and the affected KV ranges enter the \texttt{Expired} state (\S\ref{subsec:state-automaton}).

Physical translation is further gated by \emph{KV existence}: only messages that have already participated in inference have materialized token spans and KV ranges to manage. \sys therefore applies token/KV updates only at inference boundaries. Between inferences, it processes each context's events serially in program order, updating message states and recording the events against the latest materialized mapping; the next inference emits the minimal invalidation and update actions required by the new prefix structure.

\subsection{Lifecycle State Management}
\label{subsec:state-automaton}

\begin{figure}[t]
  \centering
  \includegraphics[width=0.9\linewidth]{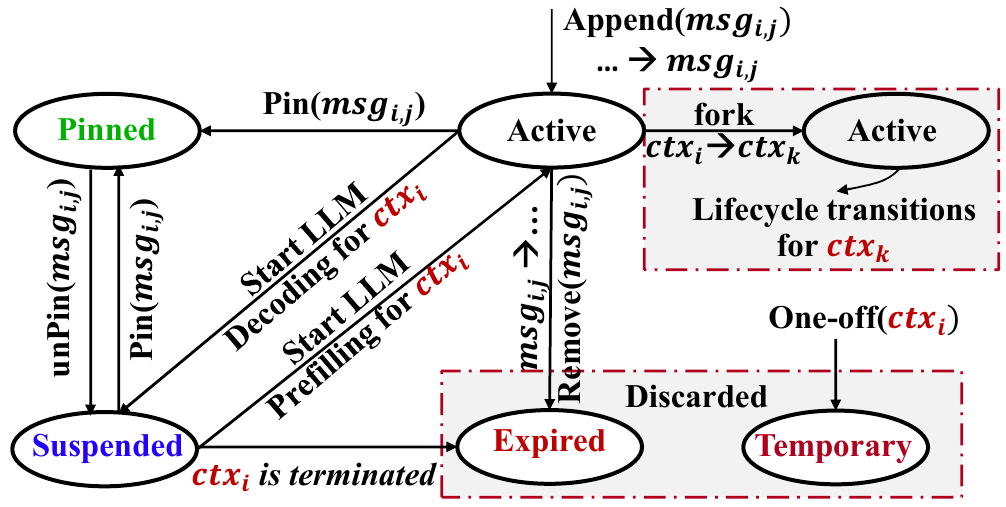}
  \caption{Message lifecycle state transitions within and across contexts
  in \sys.}
  \label{fig:lifecycle-transitions}
  \vspace{-2mm}
\end{figure}

To preserve independent retention demands, \sys maintains a lifecycle
state for each message version in each context.
Figure~\ref{fig:lifecycle-transitions} shows
transitions driven by primitive operations and runtime events.
We denote the $j$-th message in context $ctx_i$ by
$msg_{i,j}$; its version changes when its content or preceding token
prefix changes.
Four categories capture these demands. \emph{Pinned} messages
receive explicit long-term or temporary protection. Ordinary messages are
\emph{Active} when newly appended or participating in LLM requests, and
\emph{Suspended} once LLM decoding starts. \emph{Discarded} groups
\emph{Expired} messages whose lifetimes have ended and \emph{Temporary}
messages declared for one inference. Temporary messages are tracked
logically; their exclusive KV bypasses reuse storage.

\middlepara{Runtime transitions.}
When a context $ctx_i$ is spawned or forked, \sys begins monitoring its
execution and updating message lifecycle states. As messages are appended
to $ctx_i$, \sys marks each ordinary message $msg_{i,j}$ Active.
When an LLM request for $ctx_i$ completes LLM prefilling,
\sys moves all Active messages in that context to Suspended; before the
next request, it reactivates participating ordinary messages. When $ctx_i$
terminates, its ordinary messages become Expired, while Pinned messages
retain their declared protection. Releasing protection (\texttt{unPin})
returns a valid message to Suspended between calls, or to Expired if its
context has terminated.

During execution, \sys handles three further types of context changes.
(1) When $msg_{i,j}$ is removed or replaced, its old version becomes
Expired, along with old versions of subsequent messages whose token
prefixes change. Their KV is no longer reusable by $ctx_i$. \sys
tracks the replacement and affected suffix as new versions for subsequent
LLM requests.
(2) When $ctx_i$ forks a child $ctx_k$, \sys maintains separate states
for inherited messages $msg_{k,j}$ and their parent counterparts.
Parent and child states evolve independently: execution and content
changes in either context affect only its own message states.
(3) When $ctx_i$ is declared \texttt{one\_off} before inference, \sys
marks its existing and subsequently appended messages Temporary.

%% file: content/07_cache_engine.tex
\section{State-Guided KV Cache Engine}
\label{sec:data-plane}

The lifecycle manager resolves message-level semantics into KV lifecycle
states; the cache engine uses these states to make cache decisions under
uncertain reuse. Treating this problem monolithically is counterproductive:
if all KV is handled uniformly, semantically determined cases can
interfere with the decisions that should be reserved for genuinely
uncertain reuse. \sys therefore adopts a state-prioritized design that
narrows the problem in two steps. It first filters out determined cases (\S\ref{subsec:hierarchical-kv-management}), and then
focuses optimization on the residual core uncertainty among live-but-idle KV
(\S\ref{subsec:mru-suspended}).

\subsection{Separating Determined and Uncertain Reuse}
\label{subsec:hierarchical-kv-management}

\sys begins by narrowing the uncertain-reuse problem with a
state-prioritized policy. The key idea is that lifecycle state determines
the default order of cache actions during KV admission, not
merely the eviction order among live entries. When a new KV range arrives,
the cache engine consults its lifecycle state together with the states of
resident KV, and applies the decision order in
Algorithm~\ref{alg:state-priority}. This order first removes semantically
determined cases from contention, and only then lets the remaining live KV
compete for capacity.

Algorithm~\ref{alg:state-priority} makes the narrowing effect explicit.
Temporary KV is bypassed at admission, Expired KV is reclaimed before any
live eviction begins, and ordinary idle KV in the suspended state yields
capacity before stronger live states do. This order prevents semantically determined cases
from distorting eviction decisions for genuinely uncertain reuse. In
particular, it reduces the chance of wrong-direction evictions by
preventing dead KV and higher-priority live KV from distorting competition
within the uncertain idle tier.

\middlepara{Managing Shared KV.}
\sys tracks the states of all message versions mapped to each token's KV
across contexts (\S\ref{subsec:state-automaton}). Using the priorities in
Algorithm~\ref{alg:state-priority}, it assigns the KV the highest-priority
mapped state for eviction decisions and updates this aggregate whenever a mapped
version changes state. For example, if a parent removes a message but its
forked child still holds a Suspended reference, their shared KV remains
Suspended. The KV becomes Discarded only when all mapped versions are
Discarded, and can be reclaimed once any in-flight use completes.

After this filtering step, the core uncertainty is concentrated in the
suspended tier. Section~\ref{subsec:mru-suspended} futher discusses how \sys
orders this residual uncertainty.

\begin{algorithm}[t]
\caption{State-prioritized KV admission}
\label{alg:state-priority}
\footnotesize
\begin{algorithmic}[1]
\Require incoming KV $r$, resident cache $\mathcal{C}$, free capacity $F$
\Procedure{FreeByState}{$\mathcal{C}, s, \Delta$}
    \While{$\Delta > 0$ \textbf{and} $\exists$ resident KV in $\mathcal{C}$ with state $s$}
        \State release one resident KV range with state $s$
        \State update $\Delta$
    \EndWhile
\EndProcedure
\If{$state(r)=\textsc{Temporary}$}
    \State \Return \Comment{bypass Temporary KV}
\EndIf
\State $\Delta \gets size(r) - F$
\State \Call{FreeByState}{$\mathcal{C}, \textsc{Expired}, \Delta$} \Comment{reclaim Expired KV}
\State \Call{FreeByState}{$\mathcal{C}, \textsc{Suspended}, \Delta$} \Comment{evict Suspended KV}
\State \Call{FreeByState}{$\mathcal{C}, \textsc{Active}, \Delta$} \Comment{fallback to Active KV}
\If{$state(r)=\textsc{Pinned}$}
    \State \Call{FreeByState}{$\mathcal{C}, \textsc{Pinned}, \Delta$} \Comment{compete with Pinned KV}
\EndIf
\If{$\Delta \le 0$}
    \State admit $r$ into reusable storage
\EndIf
\end{algorithmic}
\end{algorithm}

\subsection{Ordering Suspended-KV Under Pressure}
\label{subsec:mru-suspended}

After the state-prioritized filtering,
the remaining uncertainty lies in the suspended tier: KV that is still
live but currently idle. The key question is therefore how to
order evictions within this tier under cache pressure. The widely adopted policy is LRU~\cite{kwon2023pagedattention, zheng2024sglang, qin2025mooncake, cheng2025lmcache}, which assumes that the most recently accessed entry is also the one closest to reuse. However, this assumption is systematically misaligned with suspended KV in ReAct agents, where \emph{newly suspended KV is often unlikely to be reused immediately} but waits during the execution of tools, sub-agents, or other physical stages.

\begin{figure}[t]
  \centering
  \includegraphics[width=0.95\columnwidth]{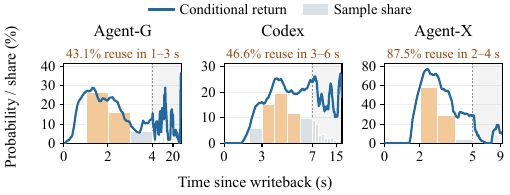}
\caption{Observed suspended KV reuse behavior of the three traces under a deployment with PD-disaggregation, where KV writeback happens after LLM prefill.}

  \label{fig:conditional-reuse-probability}
  \vspace{-1mm}
\end{figure}

Figure~\ref{fig:conditional-reuse-probability} shows this pattern. The bars show
the distribution of reuse interval since writeback, and the curve shows the conditional probability that a suspended entry will be reused within the next second given that it has remained unused up to that age. Across all three workloads, reuses are not concentrated at the youngest ages. Instead, they cluster several seconds later: in Agent-G, 43.1\% of observed returns fall within 1--3\,s after writeback; in Codex, 46.6\% fall within 3--6\,s; and in
Agent-X, 87.5\% fall within 2--4\,s.

This reversed direction determines the poor behavior of LRU under high pressure. As older suspended KV are displaced earlier than newly suspended KV, LRU tends to spend scarce residency on entries whose waiting value has not yet matured, while evicting entries that are closer to actual reuse. This effect is visible in our evaluation at the highest load points: on Agent-X, for example,
LMCache's hit ratio under LRU falls to 10.2\%, compared with 49.5\% for \sys
(\S\ref{sec:evaluation}).

Therefore, effective cache management for ReAct agents under high pressure should preserve KV long enough for its reuse value to be realized. Based on this principle, \sys follows this principle and adopts MRU (most-recently-used) as a heuristic realization: under cache
pressure, it evicts the most recently suspended KV first, favoring entries
that have waited longer and are more likely to be reused.

Note that the
advantage of this design lies in the stable directional trend in suspended-KV
reuse induced by ReAct execution, rather than precise per-entry next-use
prediction~\cite{xia2026mori, li2025continuum, tiwari2026cachewise}.
Additional queueing may lengthen reuse intervals across deployments, but the
trend remains actionable as long as the relative order of cache events is
preserved (e.g., by a first-in-first-out policy). MRU thus sustains reuse
across workloads and cache pressures (\S\ref{sec:evaluation}).

%% file: content/08_implementation.tex
\section{Implementation of \sys}
\label{sec:implementation}

\sys comprises three components. First, a lightweight in-harness framework exposes the eight primitives. It records primitive invocations synchronously to preserve the order of context mutations, and exports them asynchronously to the server-side runtime in the same order, keeping lifecycle export off the critical path without changing event order.

Second, the lifecycle manager is a server-side Python runtime that sits between the harness and the lower serving stack. It ingests lifecycle events, reproduces the LLM engine’s prompt-formatting and tokenization pipeline for independent message-to-KV translation, and interposes on LLM calls to coordinate cache actions. Before forwarding each call, it passes the call ID and context ID to the cache engine, enabling subsequent admission and bypass decisions over the correct context-owned KV ranges.

Third, \sys's state-guided policy runs on a multi-tier cache engine spanning GPU HBM, host memory, and remote memory. We separate global lifecycle control, which decides whether a page remains in reusable KV space, from per-tier caching, which decides whether a tier retains a fast copy. We provide a native engine using CUDA for local GPU-CPU movement and MoonCake for cross-node transfers, plus a compatibility path that proxies \sys's control logic onto existing backends such as LMCache and MoonCake Store.

%% file: content/09_evaluation.tex
\section{Evaluation}
\label{sec:evaluation}

In this section, we first compare \sys's end-to-end performance with
the baselines. We then conduct ablations of lifecycle tracking and recency
optimization and assess the integration effort of its primitive interface
across six agent harnesses. Finally, we measure runtime overhead.

\subsection{Experimental Setup}
\label{subsec:setup}
\begin{figure*}[!t]
  \centering
  \includegraphics[width=0.85\textwidth]{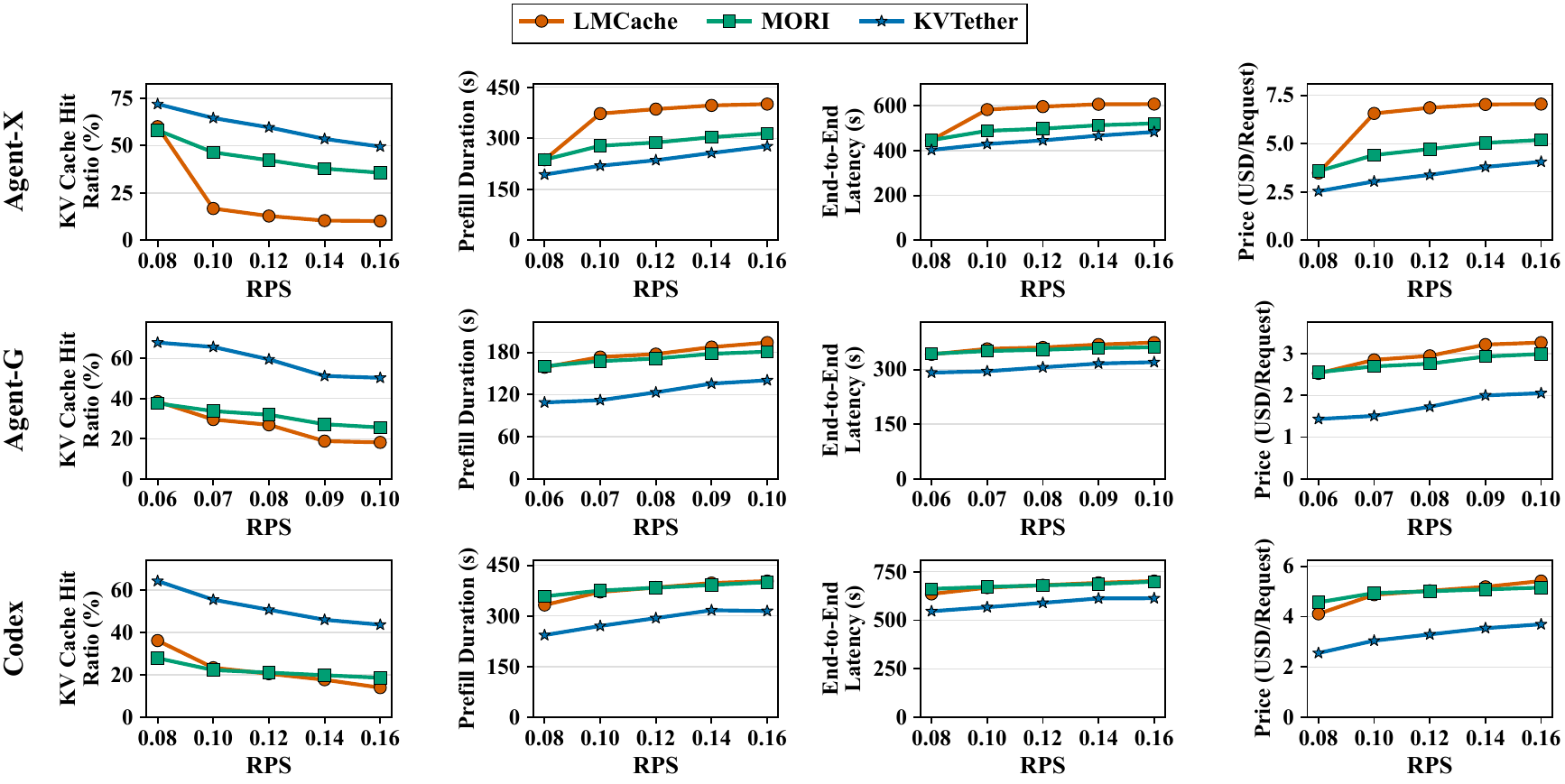}
  \caption{KV cache hit ratio (first column), cumulative prefill time
  (second column), end-to-end latency (third column), and average task
  pricing (fourth column) of \sys, LMCache, and MORI with Agent-X,
  Agent-G, and Codex.}
  \label{fig:agent-main-results}
  \vspace{-2mm}
\end{figure*}

\leadpara{Hardware and software.}
We use the three-node cluster in Table~\ref{tab:motivation-setup}, with
12 NVIDIA GB200 GPUs, six Grace CPUs, and RDMA RoCE networking.
Each node allocates 400~GB of host memory to the KV pool (1.2~TB total).
SGLang~\cite{zheng2024sglang} serves GLM-4.7~\cite{zai2025glm47}
using prefill--decode disaggregation~\cite{zhong2024distserve}, with one
prefill and two decode instances, each using tensor parallelism of four.
MoonCake~\cite{qin2025mooncake} provides cross-node
KV transfers. All methods use the same model deployment and per-tier KV
capacity.

\begin{table}[t]
    \centering
    \caption{Experimental setup.}
    \vspace{-2mm}
    \footnotesize
    \begin{tabular}{c|c}
        \hline
        & \textbf{Specification} \\ \hline
        \multirow{2}*{\textbf{Hardware}} & 3 nodes, each with 2 Nvidia Grace CPUs @ 3.38GHz \\
        & 4 Nvidia GB200 GPUs, 4$\times$200Gb/s RDMA RoCE \\
        \hline
        \multirow{2}*{\textbf{Software}} & Ubuntu 24.04, CUDA 12.9, cuDNN 9.10.2, NCCL 2.27.5 \\
        & SGLang 0.5.8, LMCache 0.5.5, MoonCake 0.3.8 \\
        \hline
        \multirow{2}*{\textbf{Model}} & GLM-4.7 (358B) with PD disaggregation \\
        & (1 prefill + 2 decode instances, TP=4) \\
        \hline
        \multirow{2}*{\textbf{Cache}} & 400~GB KV cache pool per node \\
        & KV transfer through MoonCake transfer engine \\
        \hline
        \multirow{2}*{\textbf{Dataset}} & Production trajectories of Agent-X and Agent-G \\
        & Codex \\
        \hline
    \end{tabular}
    \label{tab:motivation-setup}
\end{table}

\middlepara{Workloads.}
We replay the three workloads in \S\ref{subsec:reuse-dynamics}.
For \emph{Codex}, we collect trajectories by running the Codex agent
harness~\cite{openai2026codexsource} on
TerminalBench~\cite{merrill2026terminalbench} tasks, with all LLM requests
served by DeepSeek-V4-Pro~\cite{deepseek2026pricing}.
For \emph{Agent-X} and \emph{Agent-G}, we use execution traces from two
internal coding agents deployed by a large cloud provider. Each trace
contains 400 tasks.

For controlled comparisons of cache policies, we replay each task through LLM inference on the real cluster, using random prompts with recorded per-round input token counts and enforcing recorded output counts. We adjust prompt content across rounds to exactly match each request's potentially reusable prefix length in the trace. We simulate tool calls by waiting for their recorded durations, eliminating variability in tool execution time across experiments.

Across all three workloads, we replay lifecycle events at their original
execution points and in order relative to LLM calls. This preserves message boundaries, prefix
sharing, fork inheritance, and suffix invalidation from message
replacement or removal. For \emph{Codex}, we integrate \sys's
programming model into the harness to record events online during trace
collection. For externally collected \emph{Agent-X} and \emph{Agent-G}
traces, we reconstruct events offline from logged message updates, context
relationships, and termination records, using only information available
at that execution point. We evaluate each trace at five RPS
settings with Poisson arrivals, following prior serving
evaluations~\cite{yu2022orca,kwon2023pagedattention,zhong2024distserve}.
Task samples and arrival sequences are identical across methods.

\middlepara{Baselines.}
We compare \sys with two baselines.
\textbf{LMCache (LMC)}~\cite{cheng2025lmcache} uses LRU eviction
based on page-access recency.
\textbf{MORI}~\cite{xia2026mori} ranks agent programs by idleness,
the fraction of recent execution time spent in tool calls, including
elapsed time in an ongoing call. It favors busy programs for GPU
residency while offloading and evicting more idle programs, with admission
control at both tiers.

\middlepara{Metrics.}
We report four metrics.
\textbf{(1) KV cache utilization} is the total number of cache-hit input
tokens divided by the total number of input tokens across all rounds.
\textbf{(2) Cumulative prefill time} measures the total prefill execution
time across all rounds.
\textbf{(3) End-to-end latency} measures the elapsed time from the start
of task execution to its completion.
\textbf{(4) Average task pricing} is estimated using DeepSeek-V4-Pro's peak
rates of \$0.044, \$1.32, and \$3.96 per million cache-hit input, cache-miss
input, and output tokens, respectively~\cite{deepseek2026pricing}.
We aggregate token charges across rounds and average over tasks, applying
the same rates to all methods.

\begin{figure*}[!t]
  \centering
  \includegraphics[width=0.85\textwidth]{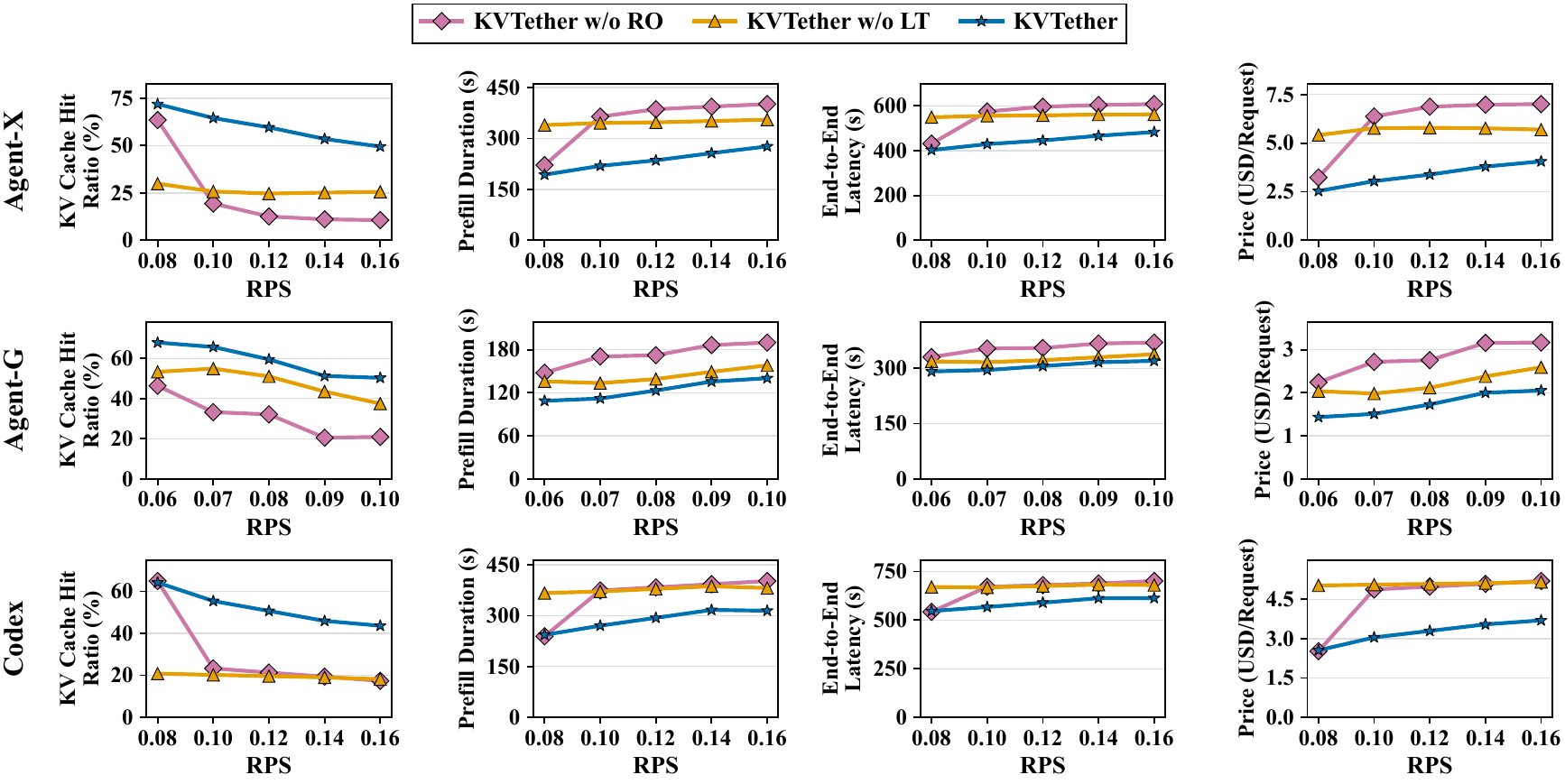}
  \caption{KV cache hit ratio (first column), cumulative prefill time
  (second column), end-to-end latency (third column), and average task
  pricing (fourth column) of \sys, \sys w/o LT, and \sys w/o RO
  with Agent-X, Agent-G, and Codex.}
  \label{fig:agent-ablation}
  \Description{KV cache hit ratio, prefill duration, end-to-end latency,
  and cost across five RPS settings, comparing \sys with variants
  without lifecycle tracking or recency optimization on three workloads.}
  \vspace{-2mm}
\end{figure*}

\subsection{End-to-End Performance}
\label{subsec:e2e-performance}

Figure~\ref{fig:agent-main-results} compares \sys with LMCache and MORI
on Agent-X, Agent-G, and Codex at five RPS settings per trace.
We report average and maximum improvements across all 15 settings.

\middlepara{KV cache hit ratio.}
\sys achieves average absolute hit-ratio gains of 33.3\% and 24.5\%
over LMCache and MORI, respectively, reaching up to 47.8\% and 36.4\%.
The gap is smaller in some low-load settings. For example, at the lowest
rate on Agent-X, LMCache and MORI retain useful KV, achieving
60.0\% and 58.1\% hit ratios, compared with \sys's 71.9\%.
At the highest rate, their hit ratios fall to 10.2\% and 35.7\%, while
\sys retains 49.5\%. \sys therefore degrades more gradually under
contention and maintains higher reuse across the evaluated load range.

\middlepara{Prefill time and end-to-end latency.}
By retaining more reusable KV under contention, \sys avoids repeated
prefix computation.
\sys reduces prefill time by 29.2\% and 23.3\% on average relative
to LMCache and MORI, respectively, and by up to 41.2\% and 33.2\%.
These savings reduce end-to-end latency by 16.5\% and 12.4\% on average
relative to LMCache and MORI, respectively, and by up to 26.3\% and 17.4\%.
The latency gains are smaller than the prefill reductions because decoding
remains necessary.

\middlepara{Average task pricing.}
\sys lowers estimated task pricing by 40.0\% and 33.2\% on average
relative to LMCache and MORI, respectively, and by up to 53.6\% and 44.1\%.
As the baselines lose reuse under load, more input tokens incur the higher
cache-miss price. \sys limits this cost increase by retaining more
cache hits, while output charges remain fixed by replay. Its reuse
advantage thus translates into cost savings across load levels.

\subsection{Ablation Study}
\label{subsec:ablation}

We compare \sys with two variants to isolate the contributions of
lifecycle tracking and recency optimization.
\emph{\sys w/o LT} disables lifecycle tracking and state-guided cache
management while retaining MRU.
\emph{\sys w/o RO} retains lifecycle tracking and state priorities but
replaces MRU with LRU within the Suspended tier.
Figure~\ref{fig:agent-ablation} reports their performance using the same
workloads, RPS settings, and cache budgets as in
Section~\ref{subsec:e2e-performance}.

\subsubsection{Lifecycle Tracking}
\label{subsec:ablation-state}

Compared with \sys w/o LT, \sys achieves an average absolute
hit-ratio gain of 25.6\% across the 15 settings, reaching up to 43.4\%.
It reduces prefill time, end-to-end latency, and average task pricing by
23.2\%, 13.1\%, and 33.1\% on average, with maximum reductions of
43.1\%, 26.6\%, and 53.2\%, respectively.
MRU alone can even underperform LMCache with LRU eviction.
For example, at the lowest RPS on Codex, \sys w/o LT achieves a 20.8\% hit
ratio, compared with 36.1\% for LMCache.

\begin{figure}[t]
  \centering
  \includegraphics[width=\columnwidth]{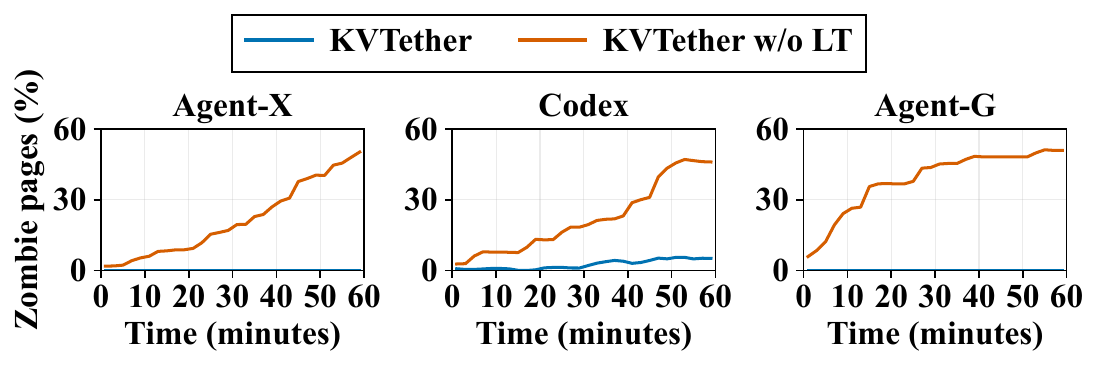}
  \caption{Zombie-page occupancy under \sys and \sys w/o LT.}
  \label{fig:ablation-zombie-pages}
\end{figure}

Figure~\ref{fig:ablation-zombie-pages} compares zombie-page occupancy
under \sys and \sys w/o LT over  one-hour windows from
workload replay. Under \sys w/o LT, zombie pages occupy an
average of 22.1\%, 21.4\%, and 38.5\% of resident KV on Codex,
Agent-X, and Agent-G, respectively. \sys exhibits only minor zombie
occupancy on Codex, averaging below 2.5\% of resident KV, and no
zombie pages on the other workloads. Without lifecycle information,
MRU can retain older zombie
pages while evicting newly arrived reusable KV. Changing eviction order
alone therefore leaves capacity occupied by KV with no future reuse.

\subsubsection{Recency Optimization}
\label{subsec:ablation-priority}

\begin{figure}[t]
  \centering
  \includegraphics[width=\columnwidth]{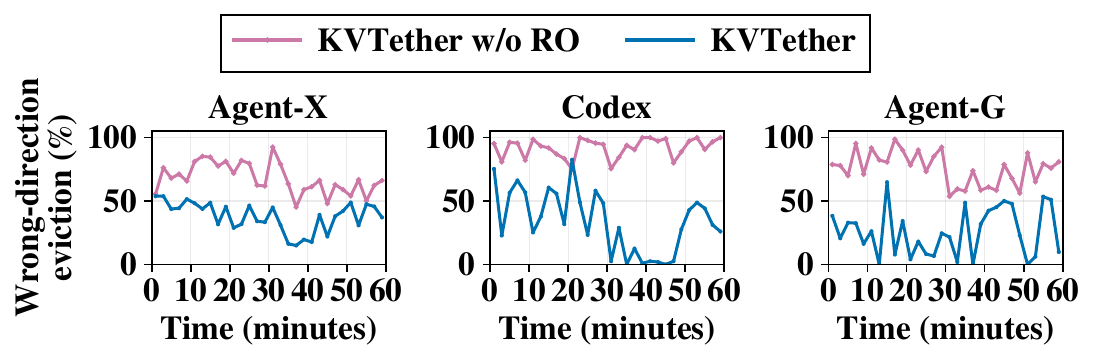}
  \caption{Wrong-direction eviction ratios under \sys and \sys
  w/o RO.}
  \label{fig:ablation-wrong-evictions}
\end{figure}

Compared with \sys w/o RO, \sys achieves an average absolute
hit-ratio gain of 29.1\%, reaching up to 47.0\%.
It reduces prefill time, end-to-end latency, and average task pricing by
26.1\%, 14.8\%, and 35.4\% on average, with maximum reductions of
39.8\%, 25.3\%, and 52.2\%, respectively.
Although \sys w/o RO reclaims obsolete KV, its LRU policy can still
evict live KV that is closer to reuse.

Figure~\ref{fig:ablation-wrong-evictions} compares wrong-direction
eviction ratios under \sys and \sys w/o RO over one-hour
windows from workload replay. Under \sys w/o RO, wrong-direction
evictions account for 65.6\%, 75.0\%, and 92.3\% of evictions on Agent-X,
Agent-G, and Codex, respectively. LRU protects newly suspended KV that may
still be waiting for tools or subagents, while evicting older KV closer
to reuse.

In contrast, \sys reduces these ratios to 38.7\%, 25.7\%, and 38.6\%,
respectively. After lifecycle filtering, MRU evicts newly suspended KV
first, reducing the risk of displacing older live KV that may be reused
earlier.

\subsection{Harness Integration}
\label{subsec:integration-effort}

This section evaluates \sys's usability by measuring the code changes
required for harness integration. We first evaluate this effort across six agent harnesses,
then use Codex as a case study to examine two representative semantic
functions, safety guardian and subagent call.

\begin{table}[t]
  \centering
  \caption{Code required to integrate \sys into six agent harnesses. ``---'' indicates that the harness does not provide
  the corresponding function.}
  \vspace{-2mm}
  \label{tab:integration-effort}
  \footnotesize
  \renewcommand{\arraystretch}{0.95}
  \setlength{\tabcolsep}{1pt}
  \begin{tabular*}{\columnwidth}{@{\extracolsep{\fill}}lrrrrrr@{}}
    \toprule
    \textbf{Component}
      & \textbf{CDX} & \textbf{SWE} & \textbf{OC}
      & \textbf{HM} & \textbf{AG} & \textbf{OCL} \\
    \midrule
    \textbf{Context manager} & 157 & 14 & 45 & 108 & 82 & 162 \\
    \midrule
    \multicolumn{7}{@{}l}{\textbf{Semantic functions}} \\
    \enspace Context append & 8 & 7 & 6 & 15 & 9 & 2 \\
    \enspace Compaction & 3 & 5 & 2 & 2 & 4 & 10 \\
    \enspace Derivation / fork & 4 & --- & 2 & 2 & 6 & 11 \\
    \enspace Safety guardian & 4 & --- & 3 & 5 & --- & 11 \\
    \enspace Deadlock detection & --- & --- & --- & --- & 10 & 10 \\
    \enspace Long-term memory & 10 & --- & --- & 2 & 6 & 10 \\
    \cmidrule(l){2-7}
    \emph{Function subtotal} & 29 & 12 & 13 & 26 & 35 & 54 \\
    \midrule
    \textbf{Total} & \textbf{186} & \textbf{26} & \textbf{58}
      & \textbf{134} & \textbf{117} & \textbf{216} \\
    \bottomrule
  \end{tabular*}
\end{table}

\middlepara{Integration effort.}
Table~\ref{tab:integration-effort} reports integration code for
Codex (CDX)~\cite{openai2026codexsource},
SWE-agent (SWE)~\cite{sweagent2026source},
OpenCode (OC)~\cite{anomaly2026opencodesource},
Hermes (HM)~\cite{nous2026hermessource}, Agent-G (AG), and
OpenClaw (OCL)~\cite{openclaw2026source}.
The counts separate reusable \emph{context-manager}
code from hooks and helpers for individual \emph{semantic functions}.
The manager encapsulates message operations and SDK calls; functions invoke
it to report context changes and retention intent. Across the six harnesses,
the manager requires 14--162 lines and function integration adds 12--54
lines, for 26--216 lines in total per harness.

\middlepara{Codex case study.}
Figure~\ref{fig:codex-integration} shows a simplified Codex integration.
Both functions reuse the wrapper in (a), which extends the native
\texttt{ContextManager} with 11 added lines to expose \sys's lifecycle
primitives.

\middlepara{Safety guardian.}
The safety guardian reviews proposed tool calls. Its ephemeral review path
invokes the manager's one-off annotation through a Session helper before
inference
(Figure~\ref{fig:codex-integration}(b)). This marks the review messages as
temporary, allowing their exclusive KV to bypass reuse storage.
The function-specific integration adds 4 lines in the example.

\middlepara{Subagent call.}
A forked subagent inherits its parent's history and executes independently.
After child creation, a Session helper registers the parent--child
relationship through the manager before task submission
(Figure~\ref{fig:codex-integration}(c)). This allows
\sys to track both contexts' lifetimes independently while preserving
shared KV needed by either context. The fork path likewise adds 4 lines
in the example.

\begin{figure}[t]
  \centering
  \setlength{\abovecaptionskip}{3pt}
  \newcommand{\rustadded}[1]{\textcolor{green!35!black}{#1}}
  \lstdefinelanguage{RustSDK}{
    keywords={use,struct,impl,fn,let,mut,for,in,if,Some,Self,async,await},
    sensitive=true,
    morecomment=[l]{//},
    morestring=[b]"
  }
  \lstset{
    language=RustSDK,
    basicstyle=\ttfamily\fontsize{7.2}{7.5}\selectfont,
    keywordstyle=\color{blue!55!black},
    commentstyle=\color{black!55},
    stringstyle=\color{green!40!black},
    moredelim=**[l][\color{green!35!black}]{+},
    moredelim=**[l][\color{red!65!black}]{-},
    columns=fullflexible,keepspaces=true,
    showstringspaces=false,breaklines=false,
    numbers=left,stepnumber=1,
    numberstyle=\tiny\color{black!45},numbersep=4pt,
    xleftmargin=12pt,framexleftmargin=12pt,
    frame=tb,framerule=0.3pt,
    rulecolor=\color{black!30},framesep=1.5pt,
    aboveskip=1pt,belowskip=0pt
  }
  \begin{minipage}{\columnwidth}
    \textbf{\footnotesize (a) Context-manager changes}
\begin{lstlisting}[firstnumber=1]
+ use (*@\rustadded{\syspkg}@*)_client::(*@\rustadded{\sysname}@*)Client; // Rust SDK
- let history = ContextManager::new();
+ let history = (*@\rustadded{\sysname}@*)ContextManager::wrap(
+   ContextManager::new(), (*@\rustadded{\sysname}@*)Client::local());
  impl (*@\sysname@*)ContextManager {
    // (*@\textcolor{black!55}{\sysname}@*) Rust SDK client stored as self.sdk.
+   fn mark_current_one_off(&self) {
+     for msg in self.messages() {
+       self.sdk.mark_one_off(&self.context_id, msg);
+     }
+   }
+   fn mark_fork_from(&self, parent_id) {
+     self.sdk.fork_context(parent_id, &self.context_id);
+   }
  }
\end{lstlisting}
    \vspace{2pt}
    \textbf{\footnotesize (b) Safety guardian (ephemeral review)}
\begin{lstlisting}[firstnumber=1]
  async fn run_ephemeral_review(...) {
    let review = ...; // Create session; prepare input.
+   review.set_(*@\rustadded{\syspkg}@*)_one_off();
    let result = run_review_on_session(&review, ...).await;
    review.shutdown_in_background();
    result
  }
+ fn set_(*@\rustadded{\syspkg}@*)_one_off(&self) { // Session method
+   self.history.mark_current_one_off();
+ }
\end{lstlisting}
    \vspace{2pt}
    \textbf{\footnotesize (c) Subagent call (fork path)}
\begin{lstlisting}[firstnumber=1]
  // spawn_agent_internal calls this, then send_input.
  async fn spawn_forked_thread(parent, ...) {
    let history = ...; // Snapshot parent history.
    let child = fork_thread_with_source(history, ...).await;
+   child.mark_(*@\rustadded{\syspkg}@*)_fork_from(parent);
    child
  }
+ fn mark_(*@\rustadded{\syspkg}@*)_fork_from(&self, parent) { // Session
+   self.history.mark_fork_from(parent.history.context_id());
+ }
\end{lstlisting}
  \end{minipage}
  \caption{Integrating \sys into Codex through context-manager extensions
  for safety guardian and subagent calls. Green \texttt{+} and red
  \texttt{-} indicate added and replaced code.}
  \label{fig:codex-integration}
  \vspace{-1mm}
\end{figure}

\subsection{Runtime Overhead}
\label{subsec:overhead}

\leadpara{Primitive recording overhead.}
\sys's in-harness framework records primitive invocations synchronously
and exports events asynchronously, adding less than 100~$\mu$s per invocation.
\textbf{LLM-call interposition overhead.}
Before forwarding each LLM call, the lifecycle manager notifies the cache
engine of the call ID and context, adding less than 50~ms.
\textbf{KV-scope resolution overhead.}
For every LLM call, the lifecycle manager replicates the inference engine's
prompt formatting and tokenization to map messages to KV ranges.
This mapping takes about 350~ms for a 256K-token input.
For inputs of this size, LLM calls within ReAct rounds in our evaluation take
19.96~s on average. The combined overhead is estimated
at 2.0\% of this duration.
\textbf{Resource overhead.}
The lifecycle runtime and cache engine's control path each run in a Python
process with one CPU core and less than 2~GB of host memory across our experiments.
To conclude, \sys's runtime and resource overheads are negligible
relative to LLM call execution time and the serving stack's resource
footprint, respectively.

%% file: content/10_conclusion.tex
\section{Conclusion}
\label{sec:conclusion}

We presented \sys, a lifecycle-aware KV cache management system for
ReAct agents that bridges the semantic gap between message-level context
management and physical KV storage. \sys exposes lifecycle semantics
through message-level primitives and translates them into KV states through
an event-driven manager that aggregates retention demands across contexts.
Guided by these states, its cache engine bypasses or reclaims
Discarded KV and applies MRU within the live-idle tier, addressing both
obsolete cache residency and premature eviction under uncertain reuse
intervals. Evaluation shows that \sys lowers estimated task pricing by
40.0\% and 33.2\% on average and end-to-end request latency by up to 26.3\%
and 17.4\% relative to LMCache and MORI, respectively. Integration into
six agent harnesses requires at most 216 lines of code per harness.